\documentclass[printer]{aa}
\usepackage{amssymb}
\usepackage{amsmath}

\usepackage{graphicx}

\newcommand{\be}{\begin{eqnarray}}
\newcommand{\ee}{\end{eqnarray}}
\newcommand{\bemul}{\begin{multline}}
\newcommand{\eemul}{\end{multline}}

\begin{document}
   \title{Magnetic field effects on neutrino production in microquasars}


  \author{M. M. Reynoso
          \inst{1,2}\fnmsep\thanks{Fellow of CONICET}
          \and
          G. E. Romero\inst{3,4}\fnmsep\thanks{Member of CONICET}
}
  \institute{Departamento de F\'{\i}sica, Facultad de Ciencias Exactas y
Naturales, Universidad Nacional de Mar del Plata,
              Funes 3350, (7600) Mar del Plata, Argentina
         \and
Instituto de Investigaciones F\'{\i}sicas de Mar del Plata, (UNMdP -
CONICET), Argentina
         \and
Instituto Argentino de Radioastronom\'{\i}a (CCT La Plata -
CONICET), C.C.5, (1894) Villa Elisa, Buenos Aires, Argentina
         \and
         Facultad de Ciencias Astron\'{o}micas y Geof\'{\i}sicas,
Universidad Nacional de La Plata, Paseo del Bosque s/n, (1900) La
Plata, Argentina}

   \date{Received September 22, 2008; accepted October 14, 2008}
\offprints{M. M. Reynoso\\
\email{mreynoso@mdp.edu.ar}}

  \abstract
{}
{We investigate the effects of magnetic fields on neutrino
production in microquasars.}
{We calculate the steady particle distributions for the pions and
muons generated in $p\gamma$ and $pp$ interactions in the jet taking
the effects of all energy losses into account.}
{The obtained neutrino emission is significantly modified due to the
synchrotron losses suffered by secondary pions and muons.}
{The estimates made for neutrino fluxes arriving on the Earth imply
that detection of high-energy neutrinos from the vicinity of the
compact object can be difficult. However, in the case of windy
microquasars, the interaction of energetic protons in the jet with
matter of dense clumps of the wind could produce detectable
neutrinos. This is because the pions and muons at larger distances
from the compact object will not be affected by synchrotron losses.}

               \titlerunning{Magnetic effects on $\nu$-production in MQs}
\maketitle

%

\section{Introduction}
Microquasars, the X-ray binary systems with non-thermal jets, are
considered important candidate sources of high-energy neutrinos
(Waxman \& Levinson 2001). The recent detection of TeV gamma rays
reveals that these objects are capable of accelerating particles to
very high energies (Aharonian et al. 2005, Albert et al. 2006,
Albert et al. 2007). The models that predict both gamma ray and
neutrino emission are based on interactions of relativistic protons
in the jet with cold protons of a dense wind from a high-mass
stellar companion (Romero et al. 2003, Christiansen et al. 2006),
with secondary synchrotron emission in the jet itself (Romero \&
Vila 2008), and with cold protons in a heavy jet (Reynoso et al.
2008). A usual assumption made in these models is equipartition
between the magnetic energy and the kinetic energy in the jets,
which leads to large magnetic fields. In this work, we analyze the
effects caused by the presence of such strong magnetic fields on the
spectra of secondary particles that decay to neutrinos.

The outline of this work is as follows. In the next section we
briefly discuss the basics of hadronic models for microquasars, and
in Sect. \ref{SecHadronic} we deal with the acceleration and cooling
mechanisms relevant to the primary relativistic particles in the
jet.  In Sect. \ref{SecMagnetic}, we analyze the effects of the
magnetic field on the spectra of secondary pions, muons, and
neutrinos. In Sect. \ref{SecClump}, we discuss the neutrino
production through interactions between the jet and clumps of the
stellar wind in high-mass microquasars. The last two sections
include a
discussion of the results and {a summary.}

\section{Basics of hadronic models of microquasars}\label{Secbasics}
\begin{figure}
\includegraphics[trim = 16mm 5mm 3mm 12mm, clip,width=9cm,angle=0]{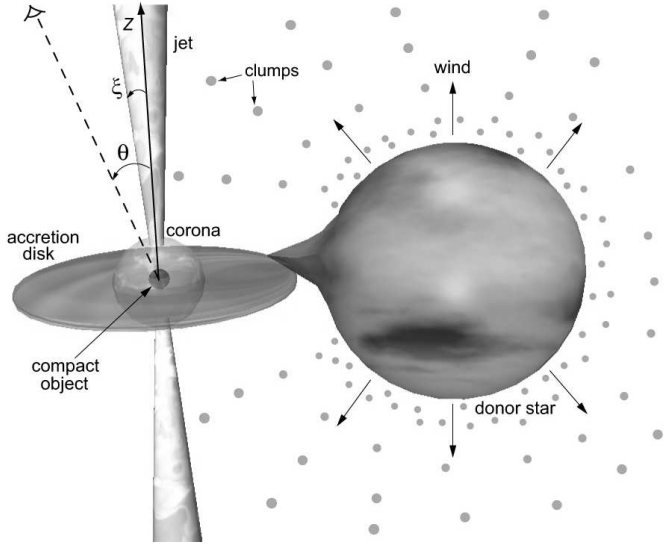}
\caption{Schematic view of a high-mass microquasar.}
\label{FigSketch}
\end{figure}

In these models, an accretion disk is present around the compact
object, and a fraction of the accreted material is expelled in two
oppositely directed jets (Falcke \& Biermann 1995). We assume
conical jets with a half-opening angle $\xi$ and radius $r(z)=
z_0\tan\xi $, where the injection point is at a distance $z_0$ from
the compact object. A sketch of a high-mass microquasar is shown in
Fig. \ref{FigSketch}, where the star presents a wind with a clumply
structure. {In the case of a low-mass microquasar, there is no
significant stellar wind.}

The kinetic luminosity of the jet, $L_{\rm k}$, implies a kinetic
energy density of \be
  \rho_k(z)= \frac{L_{\rm k}}{\pi r_{\rm j}^2 v_{\rm b}},
\ee where $v_{\rm b}$ is the bulk velocity of the jet particles.
Following the jet-accretion coupling hypothesis, we assume that
around 10\% of the Eddington luminosity goes into the jet
(K\"{o}rding et al. 2006). {We adopt $L_{\rm k}=10^{38}{\rm erg \
s}^{-1}$ for a $10 \; M_\odot$ black hole}. Equipartition then
implies a magnetic energy density $\rho_{\rm mag}= \rho_{\rm k}$,
and hence a magnetic field (e.g. Bosch-Ramon et al. 2006) \be
  B(z)= \sqrt{8\pi \rho_{\rm k}(z)}.
\ee

We consider that a fraction of the kinetic power in the jet is
carried by relativistic primary protons and electrons, $L_{\rm rel}=
L_p+ L_e$. The relation between the proton and electron power is
given by a certain parameter $a$ in such a way that $L_p= a \ L_e$.
This parameter is unknown, although there are reasons to think that
$a>1$. We consider the cases of $a=1$ for equal proton and electron
luminosities and $a=100$ for a proton-dominated jet.

\section{Hadronic processes at the base of the jet}\label{SecHadronic}

In the one-zone approximation (Khangulyan et al. 2007), we assume
that shock acceleration of the particles takes place in the jet at
distances from $z_0$ to $z_{\rm max}=5z_0$ from the compact
object.The injection rate is assumed to be a power law in the
particle energy $N'(E')=K_0 E'^{-2} {({\rm GeV}^{-1}{\rm
cm}^{-3})}$. The corresponding current can be written as $J'(E')=c
N'(E')$ in the reference frame co-moving
with the jet particles. 
The conservation of the number of particles is satisfied
if the current evolves with $z$ as (see Ghisellini et al. 1985)
 \be
   J'(E',z)={K_0 c} \left(\frac{z_0}{z}\right)^2 E'^{-2}{({\rm
 GeV}^{-1}{\rm s}^{-1}{\rm cm}^{-2})}.
  \ee

The continuity equation in the case of no time dependence and in the
absence of sinks, implies that the injection or source function of
particles must satisfy $$Q'(E',z)=\nabla\cdot J'(E',z) \; \hat{z},$$
so that we have
 \be
Q(E',z)= Q_0 \left(\frac{z_0}{z}\right)^3 E'^{-2} {({\rm
GeV}^{-1}{\rm cm}^{-3}{\rm s}^{-1})}.
 \ee
In the observer reference frame, whose line of sight makes an angle
$\theta$ with the jet, since $E'= \Gamma_{\rm b}(E- \beta_{\rm b}
\cos\theta\sqrt{E^2- m^2c^4})$, $dV'/dV= \Gamma_{\rm b}$, and
 \be
  \frac{dE'}{dE}=\Gamma_{\rm b}-\frac{\beta_{\rm b}E \cos\theta}{\sqrt{E^2-
 m^2c^4}},
 \ee
it follows that
\begin{multline}
 Q(E,z)= Q_{0}\left(\frac{z_0}{z}\right)^3{\Gamma^{-1}_{\rm b}
 \left(E-\beta_{\rm b}
 \sqrt{E^2-m^2c^4} \cos \theta \right)^{-2}}  \\
 \times \left[ \Gamma_{\rm b} -  \frac{\beta_{\rm
 b}E\cos{\theta}}{\sqrt{E^2-m^2c^4}} \right] \label{peinjection},
\end{multline}
where $\Gamma_{\rm b}$ is the bulk Lorentz factor of the jet. The
normalization constant $Q_{0}$ is obtained by specifying the power
in relativistic particles: \be
  L_{e,p}= \int_V d^3r
  \int_{E_{e,p}^{\rm(min)}}^{E_{e,p}^{\rm(max)}}dE_{e,p}
  E_{e,p}Q_{e,p}(E_{e,p},z).\label{qrelNorm}
\ee The minimum energies are $E_e^{\rm(min)}= 1 $ MeV and
$E_p^{\rm(min)}=1.2$ GeV, and the maximum energies will be obtained
in the next section by equating the acceleration rate to the energy
loss rate. The parameters of our model are summarized in Table
\ref{Tab:params}.
   \begin{table}
      \caption[]{Parameters of the model.}
         \label{Tab:params}
     $$
         \begin{array}{p{0.75\linewidth}l}
            \hline
            \noalign{\smallskip}
            Parameter      &  {\rm Value}  \\
            \noalign{\smallskip}
            \hline
            \noalign{\smallskip}
            $L_{\rm k}$: jet power &  {10^{38} {\rm erg \ s}^{-1}}   \\
            $q_{\rm rel}$: jet's content of relativistic particles & 0.1 \\
            $a$: hadron-to-lepton ratio     & 1, \  100          \\
            $z_0$: jet's launching point   & 10^8{\rm cm} \\
            $z_{\rm max}$: extent of acceleration region & 5 z_0 \\
            $\Gamma_{\rm b}$: jet's bulk Lorentz factor & 1.25 \\
            $\xi$: jet's half-opening angle & 1.5, \ 5^\circ \\
            $\theta$: viewing angle & 30^\circ \\
            $\eta$: acceleration efficiency & 0.1\\
            \noalign{\smallskip}
            \hline
         \end{array}
     $$
   \end{table}

\subsection{Accelerating and cooling rates: maximum energies}
The rate of acceleration of the particles to an energy $E$, $t_{\rm
acc}^{-1}=E^{-1}dE/dt$, is given by
 \be
t_{\rm acc}^{-1} \approx \eta \frac{c \; e \; B}{E_p}, \label{tacc}
 \ee
where we consider $\eta=0.1$ for the acceleration efficiency. {This
corresponds to the case of an efficient accelerator, as expected at
the base of the jet where shocks are mildly relativistic; see, e.g.,
Begelman et al. (1990)}.

\begin{figure*}
\begin{center}
\includegraphics[trim = 0mm 0mm 0mm 0mm, clip,
width=\linewidth,angle=0]{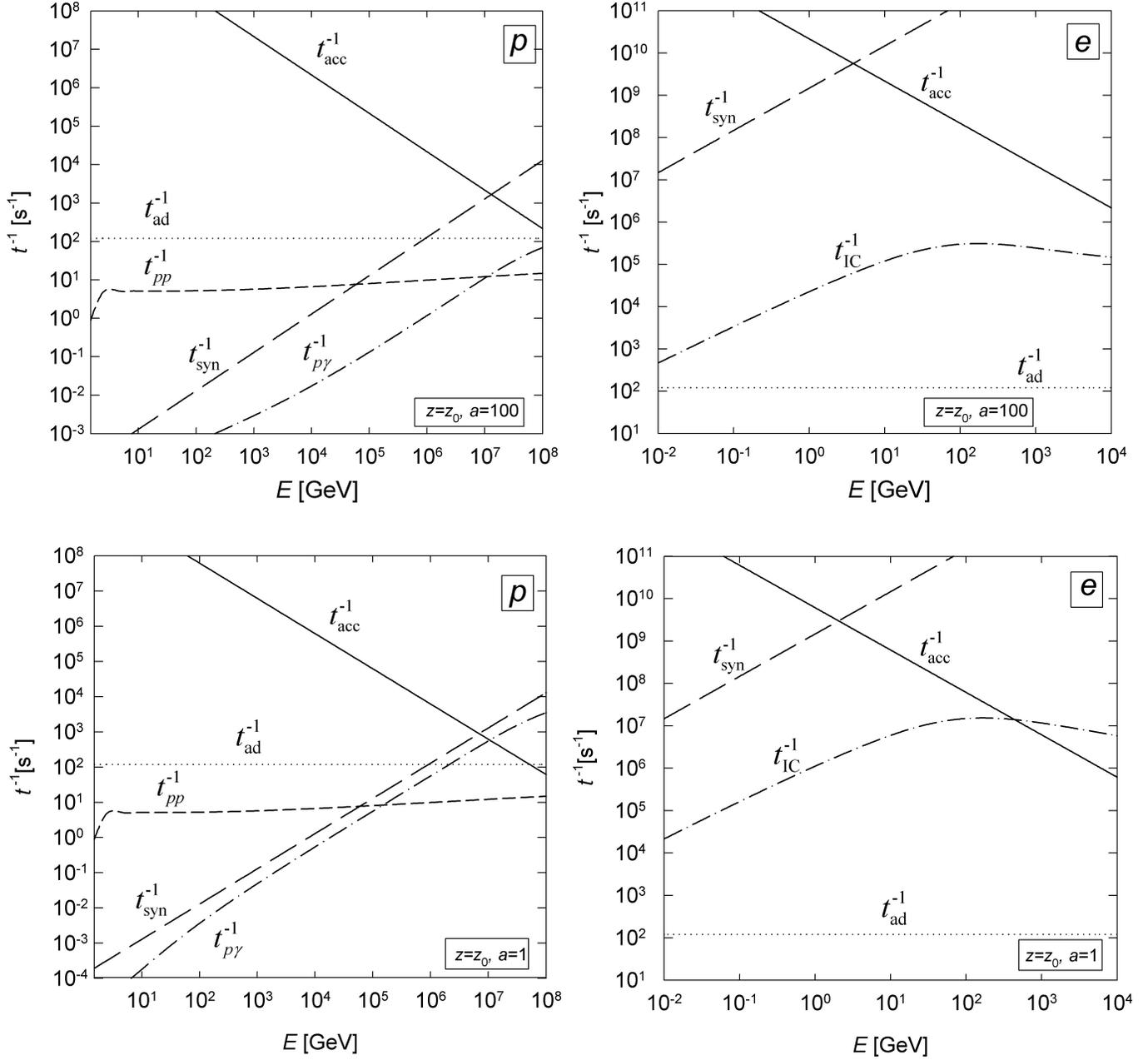}
\end{center}
\caption{{Accelerating and cooling rates for protons (left panels)
and for electrons (right panels)} at the base of the jets. The top
panels correspond to $a=100$ and the bottom ones to $a=1$. {Plots
shown: (solid lines), adiabatic cooling rates (dotted lines),
synchrotron cooling rates (long-dashed lines), $pp$ cooling rates
(short-dashed lines), $p\gamma$ cooling rates (dash-dotted lines,
left panels), and IC cooling rates (dash-dotted lines, right
panels)}.} \label{Figtpe}
\end{figure*}

Charged particles of mass $m$ and energy $E=\gamma \ m c^2$ will
emit synchrotron radiation at a rate
 \be
t_{\rm sync}^{-1}=\frac{4}{3}\left(\frac{m_e}{m}\right)^3\frac{
\sigma_{\rm T}B^2}{m_e c \ 8\pi}\gamma \label{tsyn}.
 \ee
In the jet at a distance $z$ from the compact object, the density of
cold particles is
 \be
n(z)= \frac{(1-q_{\rm rel})}{\Gamma m_p c^2 \pi r_{\rm j}^2 v_{\rm
b}}L_{\rm k}.
 \ee
The rate of $pp$ collisions of the relativistic protons with the
cold ones is then given by
 \be
t_{pp}^{-1}= n(z) \; c \; \sigma_{pp}^{\rm(inel)}(E_p)K_{pp},
 \ee
where the inelasticity coefficient is $K_{pp}\approx 1/2$ and the
corresponding cross section for inelastic $pp$ interactions can be
approximated by (Kelner et al. 2006)
\begin{multline}
\sigma_{pp}^{\rm(inel)}(E_p)= (34.3+ 1.88 L+ 0.25 L^2) \\ \times
\left[1-\left(\frac{E_{\rm th}}{E_p}\right)^4 \right]^2 \times
10^{-27}{\rm cm}^2,
\end{multline}
where $L= \ln(E_p/1000{\rm \ GeV})$ and $E_{\rm th}=1.2 {\rm \
GeV}$. Because the jet is expanding with a lateral velocity $(v_{\rm
b}\tan \xi)$ the adiabatic cooling rate is (Bosch-Ramon et al. 2006)
 \be
 t_{\rm ad}^{-1}= \frac{2}{3}\frac{v_{\rm b}}{z}.
 \ee
We estimate the maximum energies achieved by the particles by
equating $t_{\rm acc}^{-1}(E^{\rm(max)})= t_{\rm
loss}^{-1}(E^{\rm(max)})$. In the case of electrons, we assume
$t_{\rm loss}^{-1}= t_{\rm syn}^{-1}+ t_{\rm ad}^{-1}$, and for
protons $t_{\rm loss}^{-1}= t_{\rm syn}^{-1}+ t_{\rm ad}^{-1}+
t_{pp}^{-1}$. At the base of the jet, for $a=100$, we obtain \be
E_e^{\rm(max)}(z_0) \approx 7 {\ \rm GeV} \ee and \be
E_p^{\rm(max)}(z_0) \approx 10^7 {\ \rm GeV}. \ee We show in Fig.
\ref{Figtpe} the above rates for electrons and protons, as well as
the cooling rates due to $p\gamma$ and IC interactions that arise
due to the photons from the synchrotron emission (see next section).

\subsection{Proton and electron distributions}

In the one-zone approximation the particle distribution independent
of time, i.e., in a steady state, can be obtained as the solution of
the following transport equation:
 \be
 \frac{\partial{N(E,z)b(E,z)}}{\partial E}+ {t^{-1}_{\rm esc}(z)\; N(E,z)}=
 Q(E,z), \label{teq}
 \ee
where $b(E,z)= -E \; t^{-1}_{\rm loss}(E,z)$ and
  \be
  {t}^{-1}_{\rm
esc}(z)\approx \frac{c}{z_{\rm max}-z}
 \ee
 is the escape {rate}.

The corresponding solution is
\begin{multline}
N(E,z)= \frac{1}{|b(E)|}\int_E^{E^{(\rm max)}}dE'Q(E',z) \\ \times
\exp{\left\{-t^{-1}_{\rm esc}(z) \;
{\tau(E,E')}\right\}}, 
\label{Nep}
\end{multline}
with
 \be
\tau(E,E')= \int_E^{E'} \frac{dE''}{|b(E'')|}.\nonumber
 \ee
We notice that here the effect of particle acceleration is included
through the injection function which depends on the energy with a
power law in the frame co-moving with the bulk of the jet.

\begin{figure}
\includegraphics[trim = 0mm 0mm 0mm 0mm, clip,
width=8cm,angle=0]{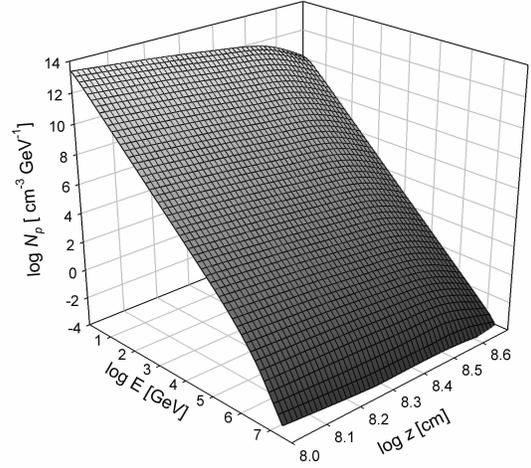}
\caption{Proton distribution as a function of energy and distance to
the compact object.} \label{FigNp}
\end{figure}

\begin{figure}
\includegraphics[trim = 0mm 0mm 0mm 0mm, clip,
width=8cm,angle=0]{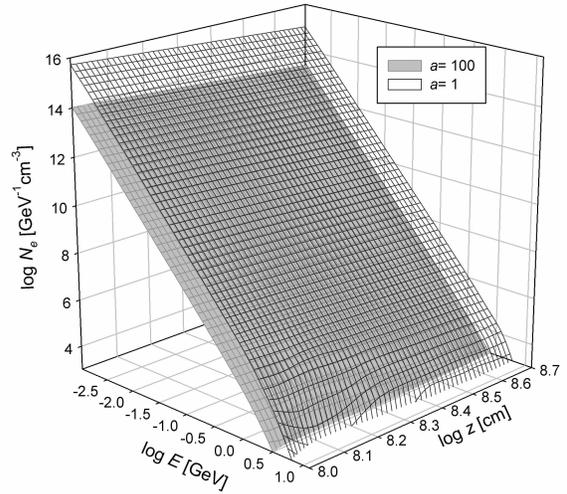}
\caption{Electron distribution as a function of energy and distance
to the compact object. The cases of $a=1$ and $a=100$ are shown {in
gray and transparent surfaces, respectively}.} \label{FigNe}
\end{figure}

We show the obtained distributions as a function of energy and $z$
in Figs. \ref{FigNp} and \ref{FigNe} for protons and electrons,
respectively. It can be seen from the latter figure that if $a=1$
the number of electrons is higher than for $a=100$, as expected. For
protons, we show the case of $a=100$, but it does not differ
significantly if $a=1$.

\subsection{Synchrotron radiation}
\begin{figure*}
\includegraphics[trim = 0mm 0mm 0mm 0mm, clip,
width=\linewidth,angle=0]{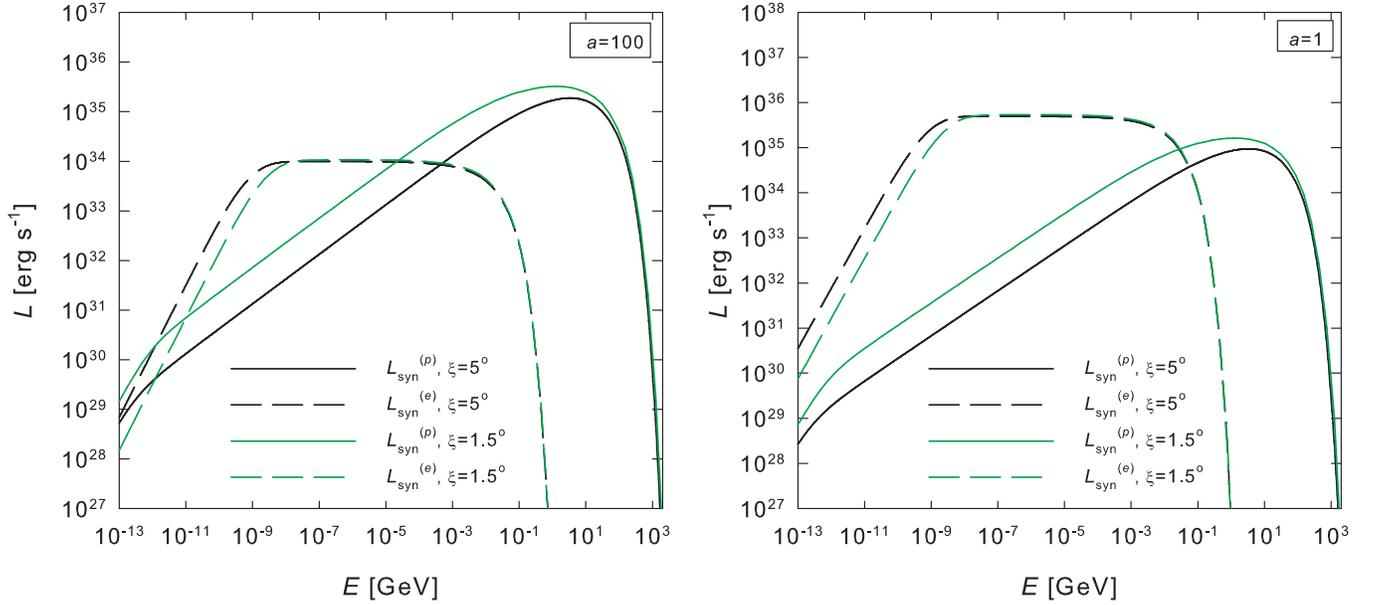}
\caption{Synchrotron luminosity emitted by protons ({solid lines})
and by electrons ({dashed lines}). Black lines correspond to
$\xi=5^\circ$ and green lines to $\xi=1.5^\circ$. The cases of $a=1$
and $a=100$ are shown in the right and left panels, respectively.}
\label{FigLsyn}
\end{figure*}

Both the protons and electrons will emit synchrotron radiation. The
power radiated by a single particle of energy $E$ and pitch angle
$\alpha$ is (e.g. Blumenthal \& Gould 1970) \be P_{\rm
syn}(E_\gamma,E,z,\alpha)= \frac{\sqrt{3} e^3 B(z)}{4\pi m c^2 h}
\frac{E_\gamma}{E_{\rm cr}} \int_{E_\gamma/E_{\rm cr}}^\infty d\zeta
K_{5/3}(\zeta), \ee where $K_{5/3}(\zeta)$ is the modified Bessel
function of order $5/3$ and
$$E_{\rm cr}= \frac{3he B(z)\sin \alpha}{4\pi m c}\left(\frac{E}{m
c^2}\right)^2.$$ The power per unit energy of the synchrotron
photons is
 \be
  \varepsilon_{\rm syn}^{(e,p)}(E_\gamma)= \int d\Omega_\alpha
  \int_{E_{e,p}^{\rm(min)}}^{E_{e,p}^{\rm(max)}} P_{\rm syn}
  N_{e,p}(E,z),
\ee and the total luminosity can be obtained by integrating in the
volume of the region of acceleration
 \be
 L_{\rm syn}^{(e,p)}(E_\gamma)= \int_V d^3 r \; E_\gamma\varepsilon_{\rm
syn}^{(e,p)}.
 \ee
The results for synchrotron radiation of protons and electrons are
shown in Fig. \ref{FigLsyn} with $a=1$ in the right panel and
$a=100$ in the left panel.

\subsection{Inverse Compton and $p\gamma$ interactions}

The synchrotron photons will, in turn, serve as targets for
electrons and protons themselves. Locally, the corresponding
radiation density can be expressed as
 \be
N_{\rm ph}(\epsilon,z)\approx \frac{\varepsilon_{\rm syn}}{\epsilon}
\frac{r_{\rm j}(z)}{c} ({\rm GeV}^{-1}{\rm cm}^{-3}).
 \ee

Electrons will interact by Inverse-Compton scatterings at a rate
  \be
t_{\rm IC}^{-1}(E,z)=\frac{4}{3}\frac{ \sigma_{\rm T}\rho_{\rm
ph}}{m_e c }\gamma_e,
 \ee
where $$\rho_{\rm ph}=\int \epsilon \;{N_{\rm ph}(\epsilon)}
d\epsilon$$ is the corresponding energy density in photons. 
As protons interact with synchrotron photons,
they lose energy due to photopion production at a rate
\begin{multline}
t_{p\gamma}^{-1}(E,z)=\frac{c}{2\gamma_p}\int_{\frac{e_{\rm
th}}{2\gamma_p}}^{\infty} d\epsilon\frac{N_{\rm
ph}(\epsilon,z)}{E_{\rm ph}^2}  \\
\times \int_{\epsilon_{\rm th}}^{2\epsilon\gamma_p} d \epsilon'
  \sigma_{p\gamma}^{(\pi)}(\epsilon')K_{p\gamma}^{(\pi)}(\epsilon')
  \; \epsilon'. \label{tpg}
\end{multline}
Here, $\epsilon_{\rm th}=150$ MeV and we adopt the cross section
(Atoyan \& Dermer 2003, see also Kelner \& Aharonian 2008)
\begin{multline}
\sigma_{p\gamma}^{(\pi)}=  \Theta({\epsilon'}-200 \ {\rm MeV}) \;
\Theta(500 \ {\rm MeV}-{\epsilon'}) \ 3.4\times 10^{-28}{\rm
cm}^2 \\
+ \Theta({\epsilon'}-500 \ {\rm MeV})\ 1.2\times 10^{-28}{\rm cm}^2,
\label{sigphotopion}
\end{multline}
and the inelasticity as
\begin{multline}
K_{p\gamma}^{(\pi)}=  \Theta({\epsilon'}-200{\rm \
MeV}) \; \Theta(500{\rm \ MeV}-{\epsilon'}) \ 0.2 \\
+ \Theta({\epsilon'}-500{\rm \ MeV})\ 0.6.
\end{multline}

The obtained IC and $p\gamma$ cooling rates are shown in Fig.
\ref{Figtpe}. It can be seen from this plot that the dominant
mechanisms for energy loss are those discussed at the beginning of
this section.

\section{Magnetic effects on secondary particles}
\label{SecMagnetic} 
\begin{figure*}
\includegraphics[trim = 0mm 0mm 0mm 0mm, clip,width=\linewidth,angle=0]{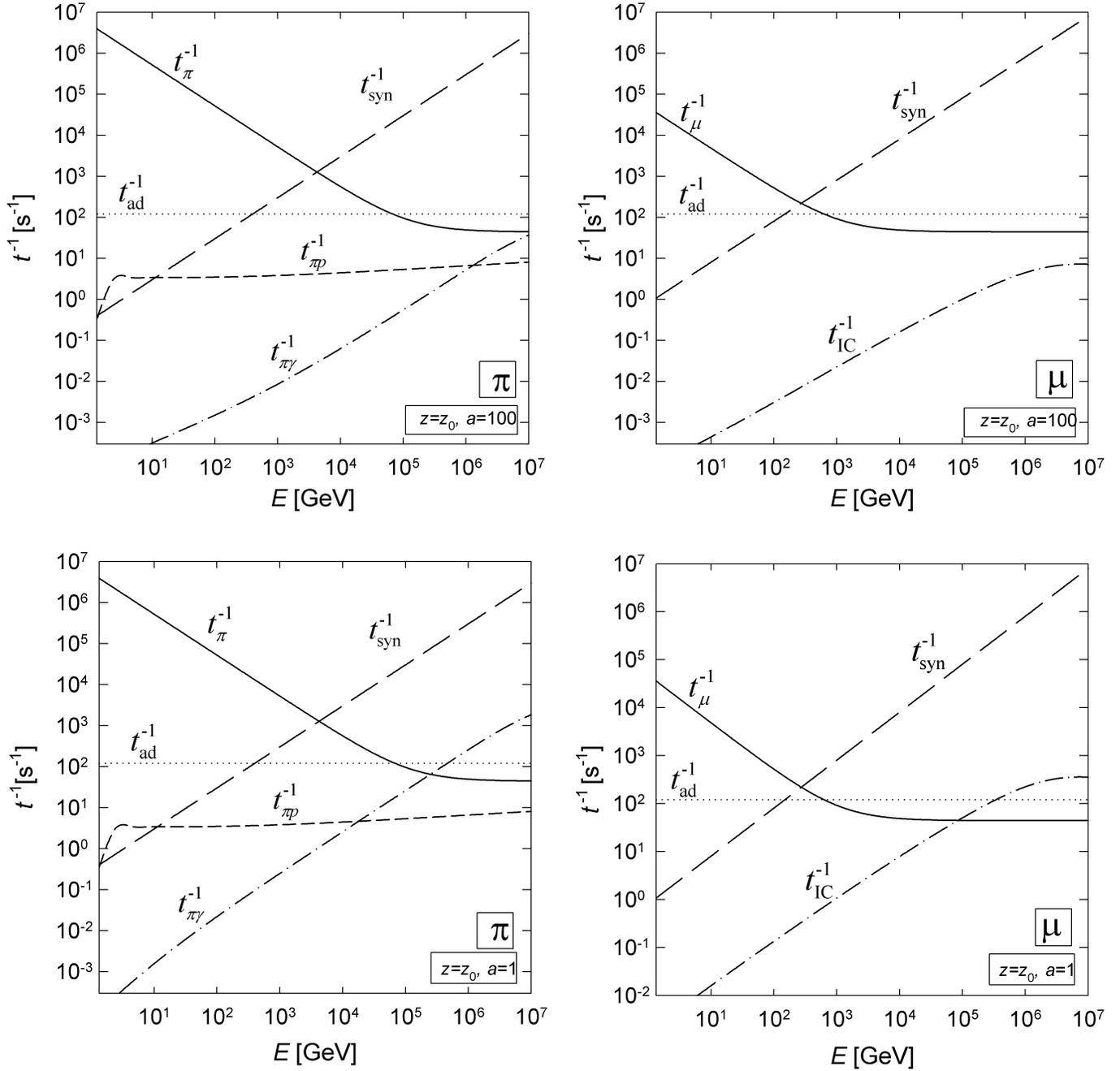}
\caption{{Cooling rates for pions (left panels) and for muons (right
panels)} at the base of the jets. The top panels correspond to
$a=100$ and the bottom ones to $a=1$. {Plots shown: adiabatic
cooling rates (dotted lines), synchrotron cooling rates (long-dashed
lines), $\pi p$ cooling rates (short-dashed lines), $\pi\gamma$
cooling rates (dash-dotted lines, left panels), and IC cooling rates
(dash-dotted lines, right panels). The decay plus escape rates are
also shown (solid lines) }}\label{Figtpimu}
\end{figure*}

The primary relativistic protons will produce
pions through inelastic interactions with matter and radiation.
Pions will decay to muons and neutrinos, and muons will also decay,
giving neutrinos and electrons:
 \be
  \pi^- \rightarrow \mu^-\bar{\nu}_\mu \rightarrow e^-\nu_\mu \bar{\nu}_e
  \bar{\nu}_\mu \\
  \pi^+ \rightarrow \mu^+ {\nu}_\mu \rightarrow e^+\bar{\nu}_\mu \nu_e
  {\nu}_\mu.
 \ee
But before decaying, pions and muons may interact, losing energy
according to the processes discussed in the previous section. In the
case of pions,
 \be b_\pi(E,z)= \frac{dE}{dt}= -E( t^{-1}_{\rm syn}+
t^{-1}_{\rm ad}+ t^{-1}_{\pi p}+ t^{-1}_{\pi\gamma}).
 \ee
For the $\pi p$ interactions we consider
 \be t^{-1}_{\pi
p}(E,z)\approx \frac{n(z) \; c \; \sigma_{\pi
p}^{\rm(inel)}(E_p)}{2},
 \ee
with $\sigma_{\pi p}(E) \approx \frac{2}{3} \sigma_{pp}^{\rm
inel}(E)$ based on the proton being formed by three valence quarks,
while the pion is formed by two (Gaisser 1990). As for the $\pi
\gamma$ interactions, we estimate a cooling rate using expression
(\ref{tpg}) with the replacement
$\sigma_{p\gamma}^{(\pi)}\rightarrow
({2}/{3})\sigma_{p\gamma}^{(\pi)} $. For muons, we have
 \be
 b_\mu(E,z)= -E( t^{-1}_{\rm syn}+ t^{-1}_{\rm
ad}+t^{-1}_{\rm IC}).
 \ee
In Fig. \ref{Figtpimu} we show the different rates corresponding to
$z=z_0$ for pions in the left panels and for muons in the right
panels. The cases with $a=100$ are shown in the upper panels and the
cases with $a=1$ in the lower panels. We have included the {rate} of
decay and escape as
 \be
t^{-1}_{\pi,\mu}(E,z)=t^{-1}_{\rm esc}(z)+ t^{-1}_{\rm dec}(E),
 \ee
where $t^{-1}_{\rm dec}=[2.6\times 10^{-8} \gamma_\pi]^{-1}({\rm
s}^{-1})$ for pions and $t^{-1}_{\rm dec}=[2.2\times 10^{-6}
\gamma_\mu]^{-1}({\rm s}^{-1})$ for muons.


\subsection{Pion injection}
The injection function of pions produced by $pp$ interactions is
given by
\begin{multline}
Q_\pi^{(pp)}(E,z)=n(z)\; c\int_{\frac{E}{E_p^{\rm
(max)}}}^{1}\frac{dx}{x}N_p\left(\frac{E}{x},z\right)
\\ \times
F_\pi^{(pp)}\left(x,\frac{E}{x}\right)\sigma_{pp}^{\rm(inel)}\left(\frac{E}{x}\right)
\label{Qpipp}
\end{multline}
where
\begin{multline}
F_\pi^{(pp)}\left(x,\frac{E}{x}\right)=4\alpha B_\pi
x^{\alpha-1}\left(\frac{1-x^\alpha}{1+r
x^\alpha(1-x^\alpha)}\right)^4 \\ \times \left(
\frac{1}{1-x^\alpha}+ \frac{r(1-2x^\alpha)}{1+
rx^\alpha(1-x^\alpha)} \right) \left(1-\frac{m_\pi c^2}{x
E_p}\right)^{1/2}
\end{multline}
is the distribution of pions produced per $pp$ collision, with $x=
E/E_p$, $ B_\pi=a'+ 0.25$, $a'= 3.67+ 0.83 L+ 0.075 L^2$, $r=
2.6/\sqrt{a'}$, and $\alpha= 0.98/\sqrt{a'}$ (see Kelner et al.
2006).

The injection function for charged pions from $p\gamma$ interactions
is
\begin{multline}
Q_\pi^{(p\gamma)}(E,z)= \int_E^{E_p^{\rm(max)}} dE_p N_p(E_p,z) \;
\omega_{p\gamma}(E_p,z) \\ \times
\mathcal{N}_\pi(E_p) \; \delta(E-0.2E_p) \\
= 5 \; N_p(5E,z) \; \omega_{p\gamma}(5E_\pi,z) \;
\mathcal{N}_\pi(5E_\pi).\label{Qpipg}
\end{multline}
Here the $p\gamma$ collision frequency is
\be \omega_{p\gamma}(E_p,z)= \frac{c}{2\gamma_p}\int_{\frac{e_{\rm
th}}{2\gamma_p}}^{\infty} d\epsilon\frac{N_{\rm
ph}(\epsilon,z)}{E_{\rm ph}^2} \int_{\epsilon_{\rm
th}}^{2\epsilon\gamma_p} d \epsilon'
  \sigma_{p\gamma}^{(\pi)}(\epsilon')
  \; \epsilon',
\ee
and the mean number of positive or negative pions is
 \be
\mathcal{N}_\pi \approx \frac{p_1}{2} + 2p_2,
 \ee
with $p_1$ and $p_2=1-p_1$ as the probabilities of single pion and
multi-pion production, respectively. These are related to the mean
inelasticity function $\bar{K}_{p\gamma}=
{t_{p\gamma}^{-1}\omega_{p\gamma}^{-1}}$ by \be
p_1=\frac{K_2-\bar{K}_{p\gamma}}{K_2-K_1}, \ee where $K_1=0.2$ and
$K_2=0.6$.

\subsection{Steady-state distribution of charged pions}

The steady pion distribution obeys the transport equation
(\ref{teq}) with the replacement $t_{\rm esc}^{-1}\rightarrow
t^{-1}_\pi(E,z)$.
%
The solution is
\begin{multline}
N_\pi(E,z)= \frac{1}{|b_\pi(E)|}\int_{E}^{E^{\rm(max)}}dE'Q(E',z)
\\ \times
\exp{\left\{-{\tau_\pi(E,E')}\right\}}.
\end{multline}
with \be \tau_\pi(E',E)= \int_{E'}^{E}
\frac{dE''t^{-1}_\pi(E,z)}{|b_\pi(E'')|}. \ee
Depending on whether we use $Q_\pi^{(pp)}(E,z)$ or
$Q_\pi^{(p\gamma)}(E,z)$ in this last expression, we obtain
$N_\pi^{(pp)}(E,z)$ or $N_\pi^{(p\gamma)}(E,z)$.

\subsection{Muon steady state distribution}

As discussed in Lipari et al. (2007), to take the muon energy loss
into account, it is necessary to consider the production of left
handed and right handed muons separately, which have different decay
spectra: \be \frac{dn_{\pi^- \rightarrow
\mu^-_L}}{dE_\mu}(E_\mu;E_\pi)= \frac{r_\pi(1-x)}{E_\pi
x(1-r_\pi)^2}\Theta(x-r_\pi) \\
\frac{dn_{\pi^- \rightarrow \mu^-_R}}{dE_\mu}(E_\mu;E_\pi)=
\frac{(x-r_\pi)}{E_\pi x(1-r_\pi)^2}\Theta(x-r_\pi), \ee with $x=
E_\mu/E_\pi$ and $r_\pi= (m_\mu/m_\pi)^2$.

The injection function of negative left handed and positive right
handed muons is
\begin{multline}
  Q_{\mu^-_L,\mu^+_R}(E_\mu,z)= \int_{E_\mu}^{E^{\rm (max)}} dE_\pi
t_{\pi,{\rm dec}}^{-1}(E_\pi)\\
\times\left(  N_{\pi^-}(E_\pi,z) \frac{dn_{\pi^- \rightarrow
  \mu^-_L}}{dE_\mu}(E_\mu;E_\pi) \right.\\ \left.
  + N_{\pi^+}(E_\pi,z)\frac{dn_{\pi^+ \rightarrow
  \mu^-_R}}{dE_\mu}(E_\mu;E_\pi) \right).
\end{multline}
Because CP invariance implies that $dn_{\pi^- \rightarrow
\mu^-_L}/dE_\mu= dn_{\pi^+ \rightarrow \mu^+_R}/dE_\mu$, and since
the above distribution obtained for all charged pions is
$N_\pi(E_\pi,z)= N_{\pi^+}(E_\pi,z)+ N_{\pi^-}(E_\pi,z)$, it follows
that
\begin{multline}
  Q_{\mu^-_L,\mu^+_R}(E_\mu,z)= \int_{E_\mu}^{E^{\rm(max)}} dE_\pi t_{\pi,
{\rm dec}}^{-1}(E_\pi)
  \\ \times
  \ N_{\pi}(E_\pi,z) \frac{dn_{\pi^- \rightarrow
  \mu^-_L}}{dE_\mu}(E_\mu;E_\pi). \label{QmuL}
\end{multline}
Similarly,
\begin{multline}
  Q_{\mu^-_R,\mu^+_L}(E_\mu,z)= \int_{E_\mu}^{E^{\rm (max)}} dE_\pi t_{\pi,
{\rm dec}}^{-1}(E_\pi)
  \\ \times
  \ N_{\pi}(E_\pi,z) \frac{dn_{\pi^- \rightarrow
  \mu^-_R}}{dE_\mu}(E_\mu;E_\pi). \label{QmuR}
\end{multline}

\begin{figure*}
\includegraphics[trim = 0mm 0mm 0mm 0mm, clip,
width=\linewidth,angle=0]{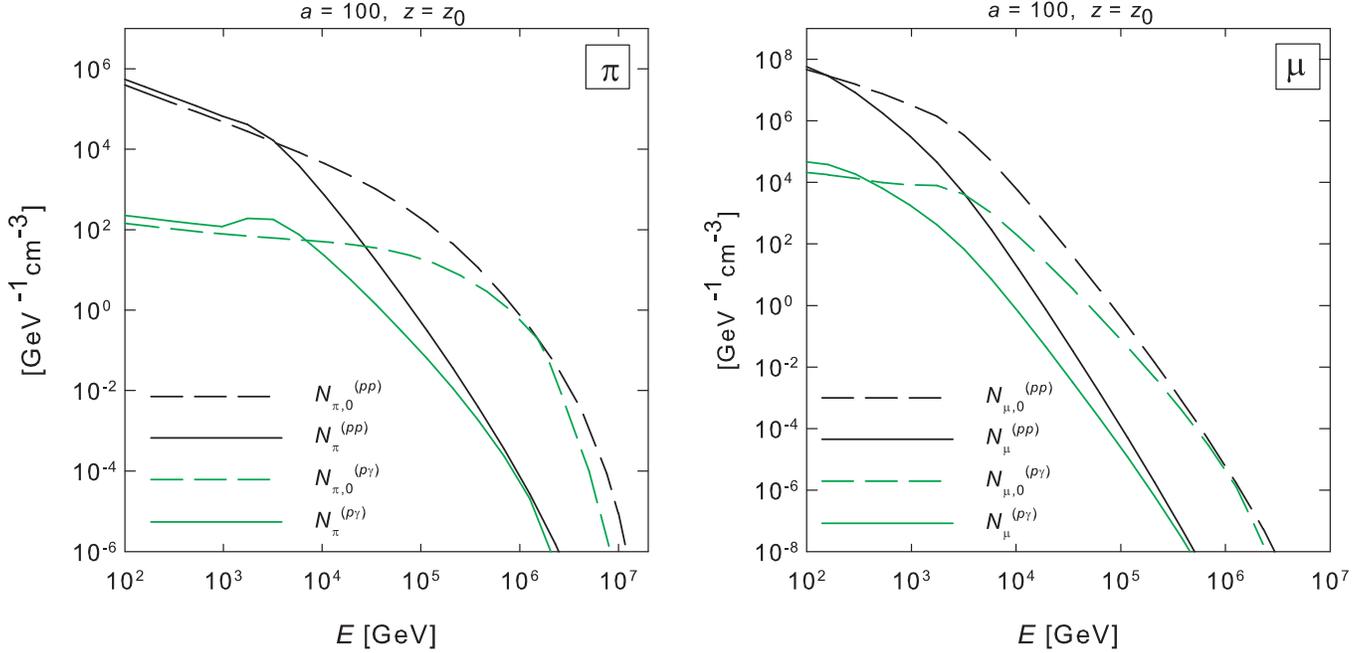}
\caption{Pion and muon distributions at the base of the jet {in the
left and right panels respectively, originated by $pp$ interactions
(black lines) and by $p\gamma$ interactions (green lines). Solid
lines: distributions obtained considering cooling. Dashed lines:
distributions obtained neglecting cooling.}} \label{FigNpimu2d}
\end{figure*}

For illustration, we show the obtained pion and muon distributions
at $z=z_0$ in Fig. \ref{FigNpimu2d}, for the cases of production
caused by $pp$ and $p\gamma$ interactions. In these plots, we also
show the particle distributions that correspond to no energy losses.
The solution corresponding to no energy losses will simply have the
form \be
   N_{\pi,0}(E,z)= \frac{Q_\pi(E,z)}{t_{\pi}^{-1}(E,z)},\\
   N_{\mu,0}(E,z)= \frac{Q_\mu(E,z)}{t_{\mu}^{-1}(E,z)}.
\ee We also note that the muon distributions shown include the
contributions of the muons with different helicity states added up.

\subsection{Neutrino emission}\label{SecNuemiss}

The total emissivity of neutrinos, $$Q_\nu(E,z)=
Q_{\pi\rightarrow\nu}(E,z)+ Q_{\mu\rightarrow\nu}(E,z),$$ is the sum
of the contribution of direct pion decays plus that of muon decays:
\begin{multline}
Q_{\pi\rightarrow\nu}(E,z)= \int_E^{E_{\rm max}}dE_\pi
t^{-1}_{\pi,\rm dec}(E_\pi)N_\pi(E_\pi,z) \\ \times
\frac{\Theta(1-r_\pi-x)}{E_\pi(1-r_\pi)},
\end{multline}
with $x=E/E_\pi$, and
\begin{multline}
Q_{\mu\rightarrow\nu}(E,z)= \sum_{i=1}^4\int_E^{E_{\rm
max}}\frac{dE_\mu}{E_\mu} t^{-1}_{\mu,\rm
dec}(E_\mu)N_{\mu_i}(E_\mu,z) \\ \times \left[\frac{5}{3}-
3x^2+\frac{4}{3}x^3+\left(3x^2-\frac{1}{3}-\frac{8x^3}{3}\right)
h_{i}\right].
\end{multline}
In this last expression, $x=E/E_\mu$,
$\mu_{\{1,2\}}=\mu^{\{-,+\}}_L$, $\mu_{\{3,4\}}=\mu^{\{-,+\}}_R$,
and
 \be 
 h_{\{1,2\}}=-h_{\{3,4\}}= -1,
 \ee 
according to Lipari et al. (2007).

\begin{figure*}
\includegraphics[trim = 0mm 0mm 0mm 0mm, clip,
width=\linewidth,angle=0]{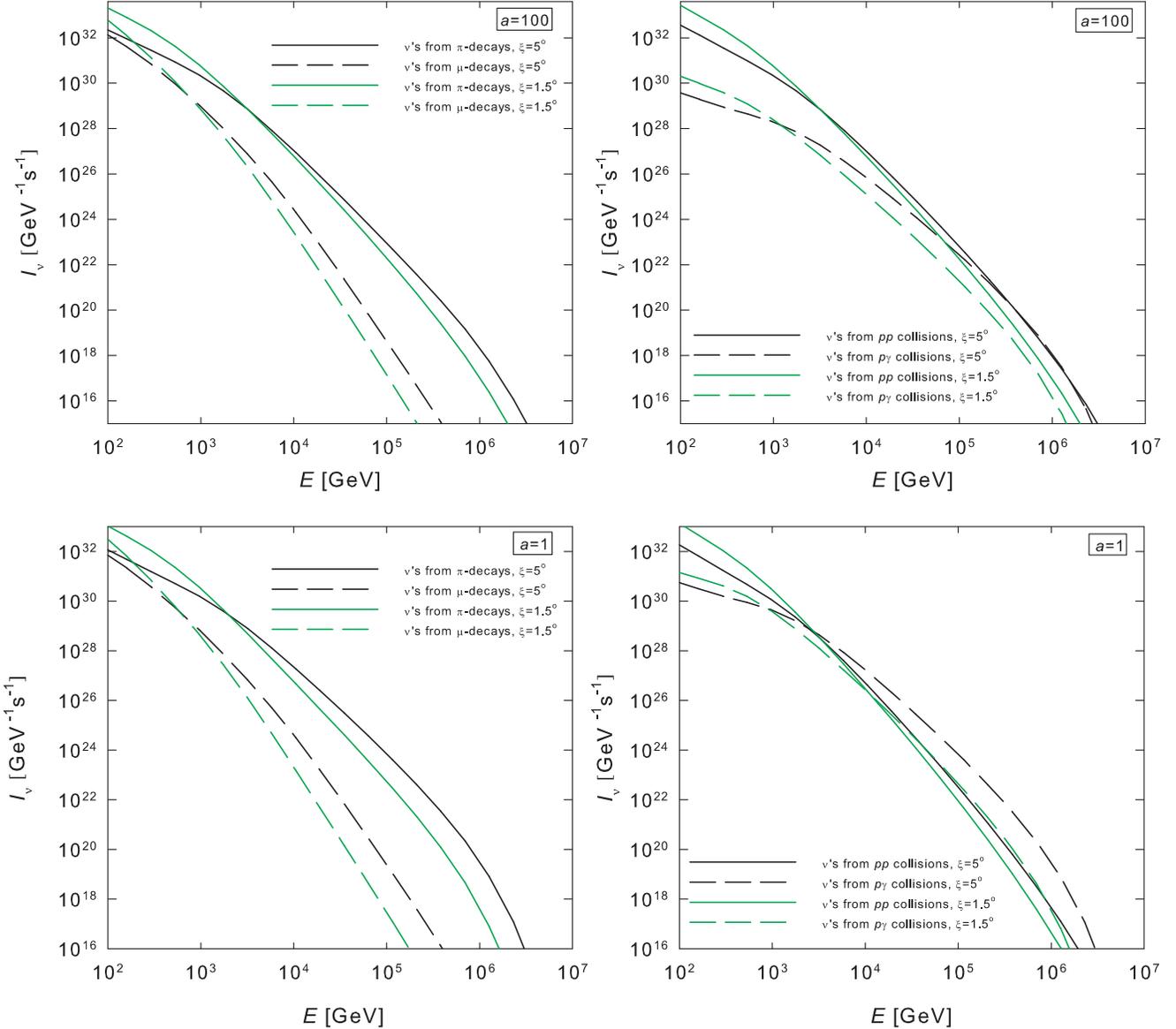}
\caption{Different contributions of the neutrino intensity produced
at the base of the jet. The cases of $a=100$ and $a=1$ are shown in
the top and bottom panels, respectively. {Black lines correspond to
$\xi=5^\circ$ and green lines to $\xi=1.5^\circ$. Left panels:
contributions from direct pion decays (solid lines) and from muon
decays (dashed lines). Right panels: contributions due to $pp$
interactions (solid lines) and due to $p\gamma$ interactions (dashed
lines).} } \label{FigInu}
\end{figure*}

The neutrino intensity (in units of ${\rm GeV}^{-1}{\rm s}^{-1}$),
 \be
I_\nu(E)= \int_V d^3 r  \; Q_\nu(E,z),
 \ee
is shown in Fig. \ref{FigInu} for the different values of $a$ and
the jet half-opening angle $\xi$.

\begin{figure*}
\includegraphics[trim = 0mm 0mm 0mm 0mm, clip,
width=\linewidth,angle=0]{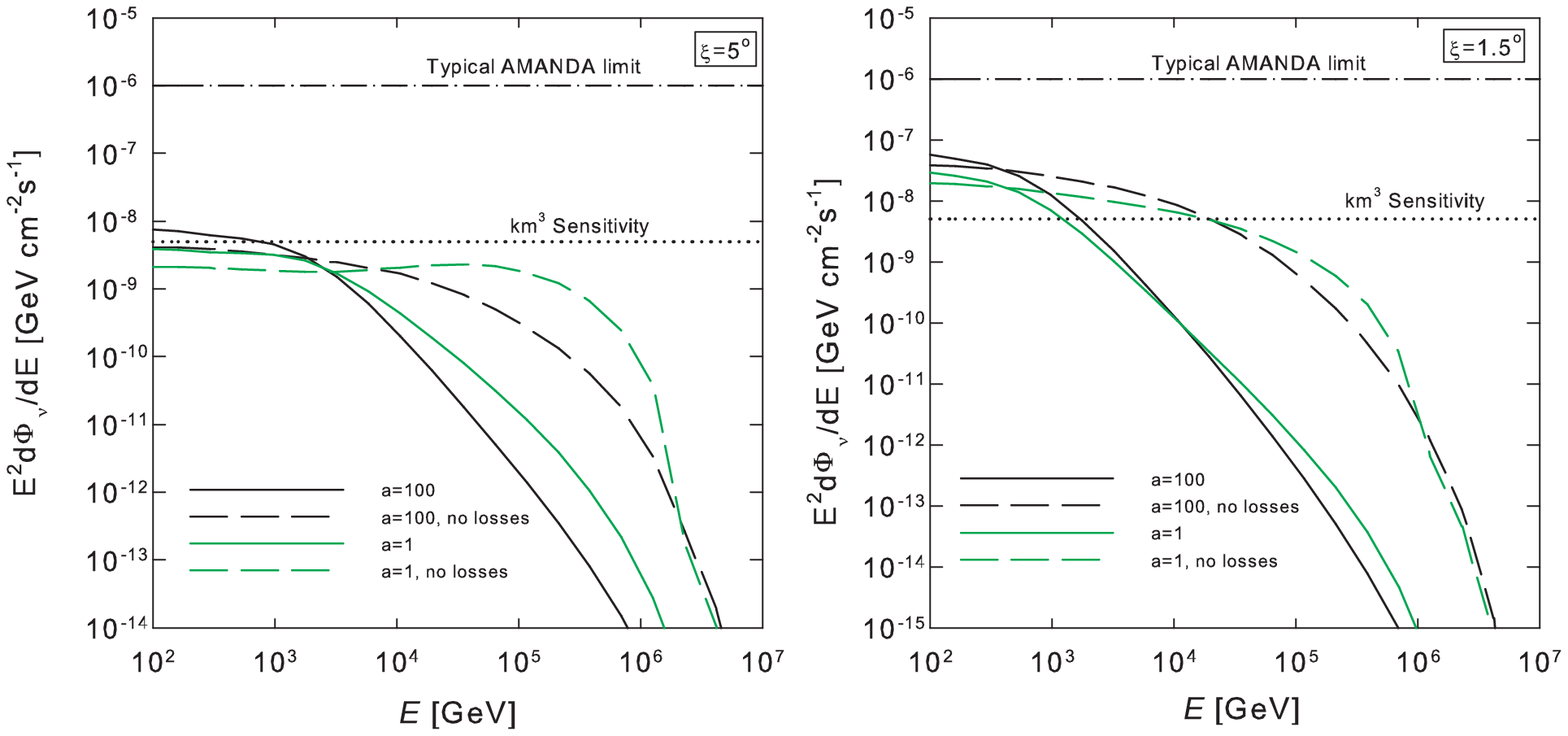}
\caption{Differential neutrino fluxes weighted by the squared
energy. The cases of $\xi=5^\circ$ and $\xi=1.5^\circ$ are shown in
the left and right panels, respectively. {Black lines correspond to
$a=100$ and green lines to $a=1$. Solid (dashed) lines: losses of
secondary pions and muons considered (neglected).} }
\label{FigE2dflu-base}
\end{figure*}

The differential flux of neutrinos arriving at the Earth can be
obtained as
 \be
\frac{d\Phi_\nu}{dE}= \frac{1}{4\pi d^2}I_\nu(E).
 \ee
This quantity, weighted by the squared energy, is shown in Fig.
\ref{FigE2dflu-base} for a source at a distance $d=2$ kpc, different
values of the jet opening angle, and different values of $a$. As a
guide, we also include a typical upper limit as derived from
AMANDA-II data, as well as the expected sensitivity for the next
generation neutrino telescope (Halzen 2006, see also Aiello et al.
2007).

\section{Neutrino production through jet-wind
interactions}\label{SecClump}

In high-mass microquasars, the donor star can present a strong wind
with dense \textsl{clumps} of matter (Romero et al. 2007). In this
section, we apply a very simplistic model to estimate the possible
high-energy neutrino emission produced by the interaction of such
clumps with jet matter.

We consider that the matter composing the clump that is successfully
interacting with the jet is uniformly distributed within a slice of
the jet with a thickness equal to the radius of the clump. This
enables us to apply the same method we used in the previous
sections, where the one-zone acceleration region was placed in the
vicinity of the compact object. In this case, the acceleration zone
will be located at high distances from the compact object, say
around half the distance to the companion star. There, the density
of jet matter is two or three orders of magnitude less than that of
the clumps, and the magnetic field is expected to be much weaker
than close to the compact star.

\begin{figure*}
\includegraphics[trim = 0mm 0mm 0mm 0mm, clip,
width=\linewidth,angle=0]{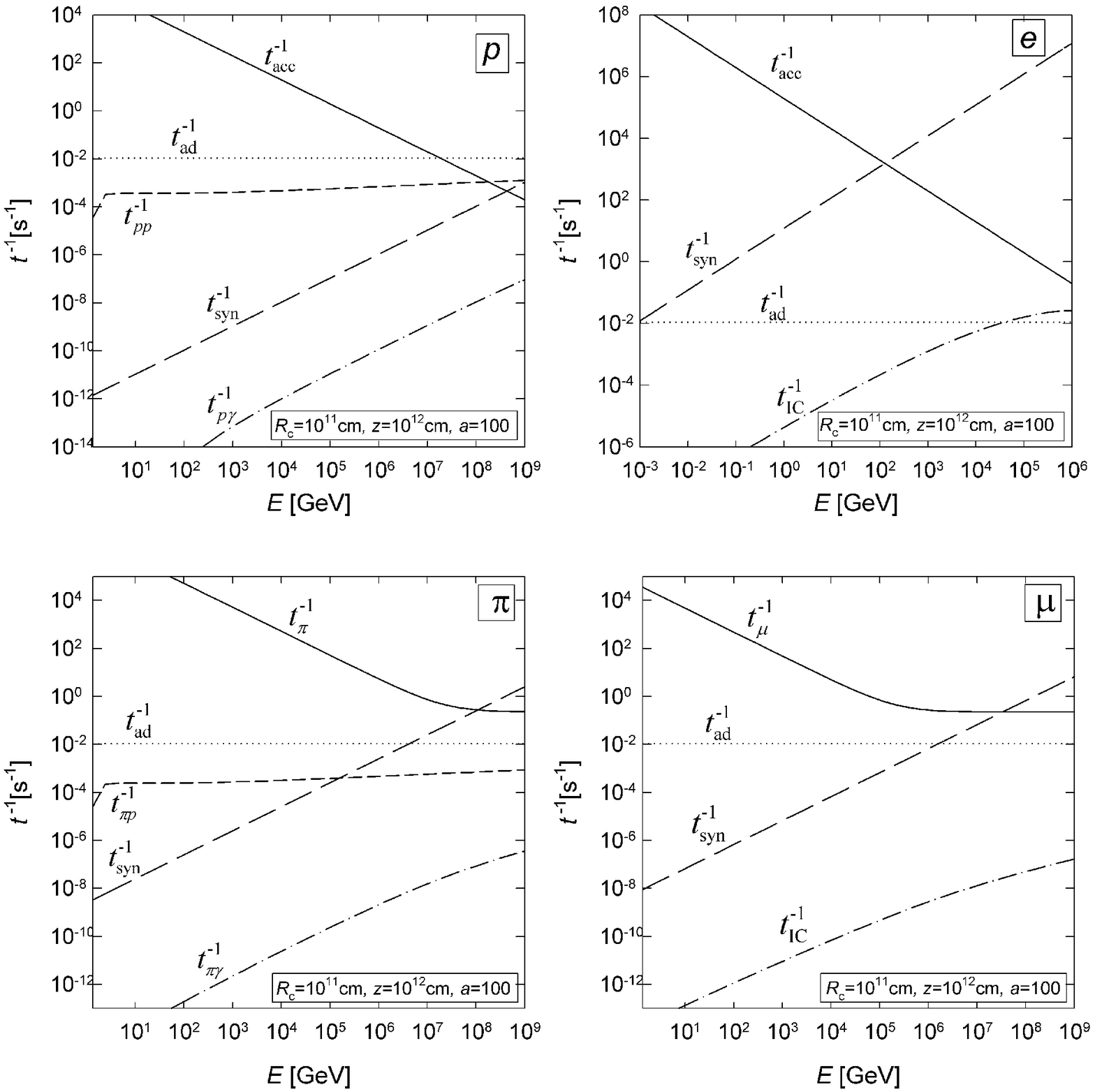}
\caption{Proton and electron accelerating and cooling rates (top
panels), and pion and muon cooling rates (bottom panels),
{for jet-clump interactions at $z=10^{12}$. Plots shown:
accelerating rate (solid lines, top panels), decay plus escape rate
(solid lines, bottom panels), adiabatic cooling rates (dotted
lines), synchrotron cooling rates (dashed lines), $pp$ cooling rate
(short-dashed line, top left panel), $\pi p$ cooling rate
(short-dashed line, bottom left panel), $p\gamma$ cooling rate
(dash-dotted line, top left panel), $\pi\gamma$ cooling rate
(dash-dotted line, bottom left panel), IC cooling rate (dash-dotted
line, right panels)} } \label{Figteppimu-cl}
\end{figure*}

We assume that the clump density is $\rho_{\rm c}=10^{-12}{\rm g \
cm}^{-3}$, and we consider two different radii: $R_{\rm
c}=10^{11}{\rm cm}$ and $R_{\rm c}= 10^{10}{\rm cm}$. The
relativistic particles are accelerated in a region of length $R_{\rm
c}$. We assume a lower acceleration efficiency, $\eta=0.01$, and we
can calculate, as above, the different cooling rates. In doing this,
we consider that the density of cold matter in the acceleration zone
is
 \be
 n_{\rm c}(z)= \frac{4}{3}\frac{R_{\rm c}^2\rho_{\rm
 c}}{r_{\rm j}^2 m_p}.
 \ee
For illustration we show in Fig. \ref{Figteppimu-cl} the obtained
acceleration and cooling rates, in the case of a successful
jet-clump interaction at $z=10^{12}$cm.

We consider that clumps can penetrate the jet at distances $\gtrsim
5\times 10^{11}$cm, where the density of cold protons in the jet
begins to decrease below  $10^9{\rm cm}^{-3}$. The protons and
electrons are then injected in a slice of thickness $R_{\rm c}$
using expression (\ref{peinjection}), and normalizing it through Eq.
(\ref{qrelNorm}) using $q_{\rm rel}=0.1$ and $a=100$. The steady
state distributions of protons and electrons are found using
expression (\ref{Nep}) taking $T_{\rm esc}= R_{\rm c}/v_{\rm b}$.
Next, the injection of pions is found to be dominated by the
contribution of $pp$ interactions, given by expression
(\ref{Qpipp}).

According to Fig. \ref{Figteppimu-cl}, the decay process of the
pions and muons dominate in almost all the relevant energy range, so
in a first approximation we can neglect losses so that
 \be
N_\pi(E,z)\approx \frac{Q_\pi^{(pp)}(E,z)}{t^{-1}_\pi(E,z)}.
 \ee
Substituting this pion distribution in expressions (\ref{QmuL}) and
(\ref{QmuR}), we can obtain
 \be N_\mu(E,z)\approx
\frac{Q_\mu^{(pp)}(E,z)}{t^{-1}_\mu(E,z)}.
 \ee

\begin{figure}[h]
\includegraphics[trim = 0mm 0mm 0mm 0mm, clip,
width=\linewidth,angle=0]{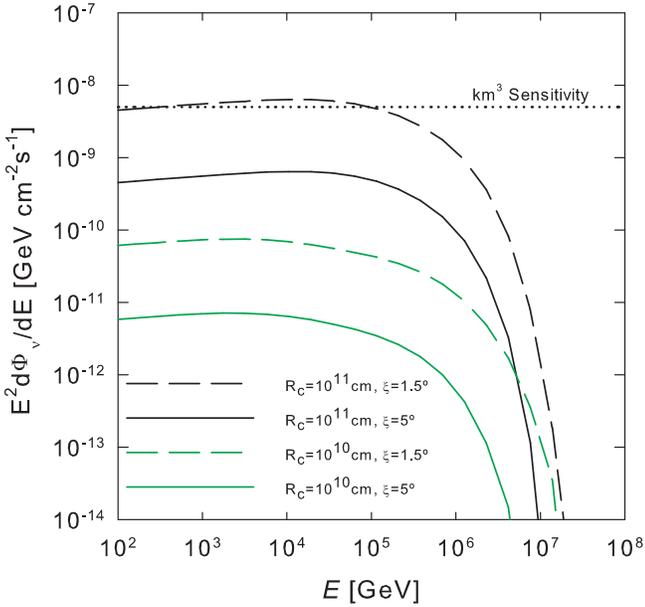}
\caption{Differential neutrino fluxes weighted by the squared
energy. The solid lines correspond to $\xi=5^\circ$ and the dashed
ones to $\xi=1.5^\circ$. The black lines correspond to $R_{\rm
c}=10^{11}$cm and the green lines to $R_{\rm c}=10^{11}$cm.}
\label{FigE2dflu-along}
\end{figure}

Following the steps discussed in Sect. \ref{SecNuemiss}, the
neutrino intensity and differential flux can be obtained. The
corresponding differential flux of neutrinos weighted by the squared
energy is shown in Fig. \ref{FigE2dflu-along}. There we show the
results corresponding to the half-opening angles $\xi=1.5^\circ$ and
$\xi=5^\circ$ for $R_{\rm c}= 10^{10}$cm and $R_{\rm c}= 10^{11}$cm.
A duty cycle of $10\%$ was adopted, meaning that the clumps interact
with the jet successfully, on average $10\%$ of the time. The
neutrinos produced are to be observed with cubic kilometer detectors
over a long period of time (several years).

\section{Discussion}

The presence of an equipartition magnetic field in the jets of
microquasars implies a strong attenuation in the high energy spectra
of pions and muons that could be produced by hadronic interactions.
This effect is found to be relevant in the vicinity of the compact
object, as can be seen in Fig. \ref{FigNpimu2d}. The neutrino flux
expected in this case is significantly reduced at energies above 1
TeV, which is the range to be probed by upcoming neutrino telescopes
such as IceCube (see Fig. \ref{FigE2dflu-base}).

Due to the lifetime of pions being shorter than that of muons,
neutrinos produced by the direct decay of pions dominate over those
originated by muon decays, since muons lose a significant fraction
of their energy by synchrotron radiation before decaying (see Fig.
\ref{FigInu}, {left panels}).


{With $a=1$, i.e. for equal power carried in relativistic protons
and electrons, the neutrino contribution due to $pp$ interactions is
dominant at low energies, $E\lesssim 10$ TeV for $\xi=1.5^\circ$ and
$E\lesssim 3$ TeV for $\xi=5^\circ$. At higher energies, the
$p\gamma$ contribution becomes important (see Fig. \ref{FigInu},
{bottom right panel}).}
However,  {in} a case with $a=100$, which seems more realistic
(Heinz 2006){, $pp$ interactions provide the most relevant mechanism
for neutrino production.}

For windy microquasars, an additional neutrino contribution can
arise from jet-wind interactions. In this case, the clumps composing
the stellar wind could interact with the jet at large distances from
the compact object ($z\gtrsim 5\times 10^{11}$cm), where the jet
particle density is much lower than that of the clumps. The magnetic
field in those regions of the jet is expected to be much lower than
at the jet base. This leads to a negligible synchrotron energy loss
of secondary pions and muons, and probably to the production of a
neutrino flux whose detectability depends on several factors such as
the density and size of the clumps, and the duty cycle corresponding
to this type of interactions.

The jet half-opening angle is another parameter that matters. We
considered two cases: $\xi= 5^\circ$ and $\xi=1.5^\circ$. The first
value is often assumed in the literature and the second one is
another possibility of a more collimated outflow (notice that for
SS433, $\xi\thickapprox 0.6^\circ$). For wide opening angles, the
magnetic energy density is lower than for narrow ones, and the
magnetic field is also lower. This leads to a lower synchrotron loss
rate and hence to a higher maximum energy of the particles. However,
for wide opening angles, the density of cold protons is lower, and
the spectrum of produced secondary particles is lower. These effects
can be seen in Figs. \ref{FigE2dflu-base} and \ref{FigE2dflu-along},
where the weighted fluxes for $\xi=5^\circ$ are lower and more
slowly decreasing with energy than those for $\xi=1.5^\circ$.

It can also be noted from these plots that, for $R_{\rm
c}=10^{11}$cm, $\xi= 1.5^\circ$, and a duty cycle of 10\%, the
expected {integrated} neutrino signal above 1 TeV from jet-clump
interactions will make the neutrino output dominate the production
at the base of the jet. In the case of low-mass microquasars;
however, only the latter will be present (Romero \& Vila 2008).

\section{Summary}

We have studied the effects caused by the magnetic field on the
secondary pions and muons that could be produced in microquasar
jets. {First, assuming an equipartition magnetic field, we
calculated the neutrino production at the base of the jet adopting
the one-zone approximation. In this case, protons and electrons are
shock-accelerated in a localized region near the compact object. A
fraction $q_{\rm rel}\sim0.1$ of the kinetic power of the jet is
transferred there to relativistic particles. The energetic protons
cool mainly by adiabatic expansion, $pp$ and synchrotron radiation.
Hot electrons cool mainly by synchrotron radiation. Secondary pions
can be produced through $pp$ and $p\gamma$ interactions, where the
targets are the cold protons and the synchrotron photons in the jet.
The pions produced also lose energy by adiabatic expansion, and
mainly through synchrotron radiation. Pions still decay giving muons
and neutrinos, but with an attenuated spectrum at high energies, due
to the effect of pion synchrotron losses. Muons also cool
significantly before decaying, giving a much lower neutrino
contribution than what was expected from previous, simpler
calculations.}

{Finally, we discussed the case of neutrino production through
jet-wind interactions in high-mass microquasars with clumpy stellar
winds. The clumps can successfully interact with the jet at large
distances from the compact object ($z\gtrsim 5\times 10^{11}$cm),
where the magnetic field is much weaker than at the jet base. We
applied a very simple model adopting the one-zone approximation at
different distances along the jet where the clumps can interact. The
size of the acceleration region is taken as the radius of the clumps
$R_{\rm c}\approx 10^{10}-10^{11}$cm, and a duty cycle of $10\%$ is
assumed. The produced pions and muons in this case do not undergo
significant cooling because the magnetic field is relatively low.
Hence, the neutrino spectrum obtained is not modified by synchrotron
radiation of the secondaries at large distances from the compact
object.}

{As a conclusion, we find that} the main contribution to the
neutrino {emission} stems from $pp$ interactions. Pions and muons
{produced in the vicinity of the compact object} are strongly
affected by synchrotron losses, and their spectra are attenuated at
high energies. As a consequence of that, we find that the neutrino
flux is much less than expected when these effects are not taken
into account. {An additional neutrino contribution, arising from
jet-wind interactions in high-mass microquasars, is not
affected by these magnetic effects. Still,} 
the detection of a neutrino signal from microquasars {seems}
difficult, but not impossible with next-generation neutrino
telescopes such as IceCube, depending on the specific parameters of
the system.

The simple models presented here serve to illustrate these effects,
which are crucial for assessing the detectability of a neutrino
signal from this type of sources. A more realistic {treatment}
including the convection of particles in the jets and a consistent
description of the acceleration mechanism will be presented
elsewhere.

\begin{acknowledgements}
We are very grateful to Gabriela S. Vila for fruitful discussions on
the topics of this paper. We also thank Prof. Hugo R. Christiansen
for useful comments on the particle injection. M.M.R. is supported
by CONICET, Argentina and Universidad Nacional de Mar del Plata
(Argentina). G.E.R. is supported by the Argentine agencies CONICET
(PIP 452 5375) and ANPCyT (PICT 03-13291 BID 1728/OC-AR). G.E.R. is
also supported by the Ministerio de Educaci\'{o}n y Ciencia (Spain)
under grant AYA 2007-68034-C03-01, FEDER funds.
\end{acknowledgements}

\end{document}